\documentclass[aps,prb,10pt,twocolumn,longbibliography,superscriptaddress]{revtex4-2}
\usepackage{amsmath}
\usepackage{amsfonts}
\usepackage{amssymb}
\usepackage{graphicx}
\usepackage{color}
\usepackage{hyperref}
\hypersetup{colorlinks=true,allcolors=blue}

\newcommand{\bq}{\mathbf{q}}
\newcommand{\sbt}[1]{_\text{#1}} 
\newcommand{\spt}[1]{^\text{#1}} 
\newcommand{\BRA}[1]{\langle #1\vert} 
\newcommand{\KET}[1]{\vert #1\rangle} 
\newcommand{\OP}[2]{\KET{#1}\BRA{#2}} 
\newcommand{\PRO}[1]{\OP{#1}{#1}} 
\newcommand{\ABS}[1]{\vert #1\vert} 
\newcommand{\dt}{\mathrm{d}t\,} 
\newcommand{\ddt}{\frac{\mathrm{d}}{\mathrm{d}t}} 
\newcommand{\LB}[2]{\mathcal{L}_{#1,#2}} 
\newcommand{\Tr}[2]{\mathrm{Tr}\sbt{#1}\left\lbrace #2 \right\rbrace} 
\newcommand{\sH}{\hat{\sigma}\sbt{H}}
\newcommand{\sV}{\hat{\sigma}\sbt{V}}

\newcommand{\FWHM}{\Delta\sbt{FWHM}}
\newcommand{\rp}{\rho\spt{2p}}
\newcommand{\rpe}[4]{\rp_{#1 #2, #3 #4}}
\newcommand{\rpH}{\rpe{H}{H}{H}{H}}

\newcommand{\rpHV}{\rpe{H}{V}{H}{V}}
\newcommand{\rpVH}{\rpe{V}{H}{V}{H}}
\newcommand{\rpc}{\rpe{H}{H}{V}{V}}
\newcommand{\rpj}{\rp_{jk,\ell m}}
\newcommand{\G}{G^{(2)}}
\newcommand{\Gj}{\G_{jk,\ell m}}
\newcommand{\Ga}{\overline{G}^{(2)}}
\newcommand{\Gaj}{\Ga_{jk,\ell m}}

\begin{document}

\title{Two-photon excitation with finite pulses unlocks pure dephasing-induced degradation of entangled photons emitted by quantum dots}

\author{T. Seidelmann}
\email[Corresponding author: ]{tim.seidelmann@uni-bayreuth.de}
\affiliation{Lehrstuhl f{\"u}r Theoretische Physik III, Universit{\"a}t Bayreuth, 95440 Bayreuth, Germany}
\author{T. K. Bracht}
\affiliation{Institut f{\"u}r Festk{\"o}rpertheorie, Universit{\"a}t M{\"u}nster, 48149 M{\"u}nster, Germany}
\affiliation{Condensed Matter Theory, Department of Physics, TU Dortmund, 44221 Dortmund, Germany}
\author{B. U. Lehner}
\affiliation{Institute of Semiconductor and Solid State Physics, Johannes Kepler University Linz, 4040 Linz, Austria}
\affiliation{Secure and Correct Systems Lab, Linz Institute of Technology, 4040 Linz, Austria}
\author{C. Schimpf}
\affiliation{Institute of Semiconductor and Solid State Physics, Johannes Kepler University Linz, 4040 Linz, Austria}
\author{M. Cosacchi}
\affiliation{Lehrstuhl f{\"u}r Theoretische Physik III, Universit{\"a}t Bayreuth, 95440 Bayreuth, Germany}
\author{M. Cygorek}
\affiliation{Heriot-Watt University, Edinburgh EH14 4AS, United Kingdom}
\author{A. Vagov}
\affiliation{Lehrstuhl f{\"u}r Theoretische Physik III, Universit{\"a}t Bayreuth, 95440 Bayreuth, Germany}
\author{A. Rastelli}
\affiliation{Institute of Semiconductor and Solid State Physics, Johannes Kepler University Linz, 4040 Linz, Austria}
\author{D. E. Reiter}
\affiliation{Condensed Matter Theory, Department of Physics, TU Dortmund, 44221 Dortmund, Germany}
\author{V. M. Axt}
\affiliation{Lehrstuhl f{\"u}r Theoretische Physik III, Universit{\"a}t Bayreuth, 95440 Bayreuth, Germany}

\begin{abstract}
  Semiconductor quantum dots have emerged as an especially
promising platform for the generation of polarization-entangled photon
pairs. However, it was demonstrated recently that the two-photon excitation
scheme employed in state-of-the-art experiments limits the achievable degree of
entanglement by introducing which-path information.  In this work, the combined
impact of two-photon excitation and longitudinal acoustic phonons on photon
pairs emitted by strongly-confining quantum dots is investigated. It is found
that phonons further reduce the achievable degree of entanglement even in the limit
of vanishing temperature due to
phonon-induced pure dephasing and phonon-assisted one-photon processes, which
increase the reexcitation probability. In addition, the degree of
entanglement, as measured by the concurrence, decreases with rising temperature
and/or pulse duration, even if the excitonic fine-structure splitting is absent
and when higher electronic states are out of reach. Furthermore, in the case of
finite fine-structure splittings, phonons enlarge the discrepancy in concurrence
for different laser polarizations.   
\end{abstract}

\maketitle

\section{Introduction}
\label{sec:intro}

In the past decades, many fascinating applications emerged in novel quantum
technologies, such as quantum
cryptography \cite{Gisin:02,Lo_quantum_cryptography,Christian_Schimpf_Crypto_2021,Basset_quantum_key_2021},
quantum communication \cite{duan_quantum_comm}, or quantum information technology
and computing \cite{pan:12,Bennett:00,Kuhn:16,Zeilinger_entangled}, that are
based on the concept of entanglement, a genuine quantum phenomenon. Especially
entangled photons attracted a lot of interest as photons travel at the speed of
light and are hardly influenced by their
environment \cite{Orieux_entangled}. Semiconductor quantum dots (QDs) show the
potential to be an excellent on-demand source of high-quality
polarization-entangled photon pairs, as demonstrated in various
experiments \cite{Benson_2000_QD_cav_device,Stevenson2006,Young_2006,Hafenbrak,Muller_2009,dousse:10,Biexc_FSS_electrical_control_Bennett,stevenson:2012,entangled-photon1,Trotta_highly_entangled_2014,winik:2017,huber2017,Huber_PRL_2018,Wang_2019,Liu2019,Hopfmann_2021,
Fognini_2019,Bounouar18,Huber_overview_2018} and theoretical
studies \cite{BiexcCasc_Carmele,Different-Concurrences:18,Seidelmann2019,Jahnke2012,heinze17,Phon_enhanced_entanglement,EdV,Troiani2006}.

Focusing on its biexciton-exciton cascade, a QD represents a quantum emitter
with four levels in a diamond configuration (cf. Fig.~\ref{fig:system}), that,
in principle, is able to generate maximally entangled photon pairs. After the
preparation of the biexciton state, two photons are emitted while the QD relaxes
back into its ground state. Because there exist two different exciton states,
this relaxation can take two different paths, resulting in the creation of
either two horizontally ($H$) or two vertically ($V$) polarized photons. Thus,
in an ideal situation, i.e., when both decay paths are otherwise
indistinguishable, a maximally entangled two-photon state
\begin{equation}
\label{eq:definition_Phi+}
\KET{\Phi_+} = \left( \KET{HH} + \KET{VV} \right) / \sqrt{2}
\end{equation}
is generated. However, in typical QDs, the exchange interaction
causes the two exciton states to be energetically split by the so-called
fine-structure splitting (FSS). Due to the FSS, the emitted photons can be
distinguished by their energies. Thus, the FSS introduces which-path information
into the system as both decay paths are no longer indistinguishable, resulting
in a reduced degree of entanglement. In the measured two-photon state, this
which-path information manifests itself as a reduced coherence between the
states $\KET{HH}$ and $\KET{VV}$ caused by a time-dependent phase oscillation
between the exciton states with a frequency proportional to the magnitude of the
FSS \cite{Hudson2007}.

Several approaches to eliminate the FSS have been
pursued \cite{Muller_2009,Biexc_FSS_electrical_control_Bennett,huber2017,
Huber_PRL_2018} and state-of-the-art experiments employing resonant two-photon
excitation (TPE) have reported unprecedented degrees of
entanglement \cite{winik:2017,huber2017,Huber_PRL_2018,Wang_2019,Liu2019,Hopfmann_2021,
Christian_Schimpf_Crypto_2021,Christian_Schimpf_strongly_ent_2021}. However, a
perfectly entangled two-photon state $\KET{\Phi_+}$ has not yet been
observed. Rather, it was shown recently that the TPE scheme itself limits the
achievable degree of entanglement by introducing an energy splitting between the
exciton states via the AC Stark effect, and thus which-path information, during
the duration of the excitation
pulse \cite{Seidelmann_PRL_2022,BassoBasset_TPE_Limit_exp_2022_arXiv}.

In contrast to other possible realizations of a four-level quantum emitter, like
F-centers or atoms \cite{edamatsu2007entangled,Freedman_FLE_atom,Park_FLE_atom},
QDs unavoidably interact with their semiconductor environment, giving rise to
electron-phonon interactions. The pure dephasing-type coupling to longitudinal
acoustic (LA) phonons in strongly-confining QDs was shown to impact the
entanglement of emitted photon pairs in the case of finite which-path
information. Depending on other system parameters, the phonon impact can range
from phonon-induced enhancement \cite{Phon_enhanced_entanglement} over a
partially reduced degree of entanglement \cite{Seidelmann2019} to a complete loss
of entanglement, which takes place for strong constant
excitation \cite{Seidelmann_Transition_2022}. Note that for completely
symmetric level structures without FSS, pure dephasing does not reduce the
degree of entanglement when an initially prepared biexciton is
assumed \cite{Seidelmann2019,BiexcCasc_Carmele,Different-Concurrences:18}. However,
this may change when the finite time of the excitation process is explicitly
taken into account, especially because the TPE scheme itself introduces
which-path information. In particular, it has previously been demonstrated that
in the case of a finite FSS phonons can reduce the degree of entanglement by
increasing dephasing and assisting off-resonant single-photon
transitions \cite{Seidelmann2019}.
In view of all the above-mentioned different ways how phonons may affect the
degree of entanglement generated in an exciton-biexciton system,
it is unclear -- without further
investigations -- what role phonons may play when combined with
the effects introduced by the finite pulse duration needed for
TPE preparation of the system in the biexciton state.

In this work, we investigate the combined impact of the TPE scheme and LA
phonons on polarization-entangled photon pairs emitted by strongly-confining
GaAs-based quantum dots. Section~\ref{sec:theory} specifies the theoretical
model for the optically excited QD as well as the calculation scheme for the
two-photon density matrix and the degree of entanglement. Based on this model,
the degree of entanglement, as measured by the concurrence, is studied as a
function of temperature and pulse duration in Section~\ref{sec:results}. Indeed,
it is found that the combination of TPE and the coupling to LA phonons results
in an additional decrease of the concurrence. For practical applications it is
equally important to know how this additional drop depends on temperature.
This is not only relevant at cryogenic temperatures that are usually used
in attempts to maximize the degree of entanglement. While at such low
temperatures one usually wants to know what limit is set by phonons to
the maximum level of entanglement, at higher temperatures the focus shifts to the question of
whether there is still enough entanglement left to realize a useful source of entangled
photons at temperatures that are more convenient for practicable applications.
We have therefore determined the temperature dependence of the concurrence
up to more elevated temperatures and find
a decreasing concurrence with rising temperature, even if the FSS
vanishes. Furthermore, this detrimental effect increases with rising pulse
duration. However, we find that at 50 K the emitted photons can still be entangled
to 90\% indicating promising perspectives for applications at
elevated temperatures.

\section{Theoretical model and evaluation of entanglement}
\label{sec:theory}

\begin{figure}[t]
\centering
\includegraphics[width=\columnwidth]{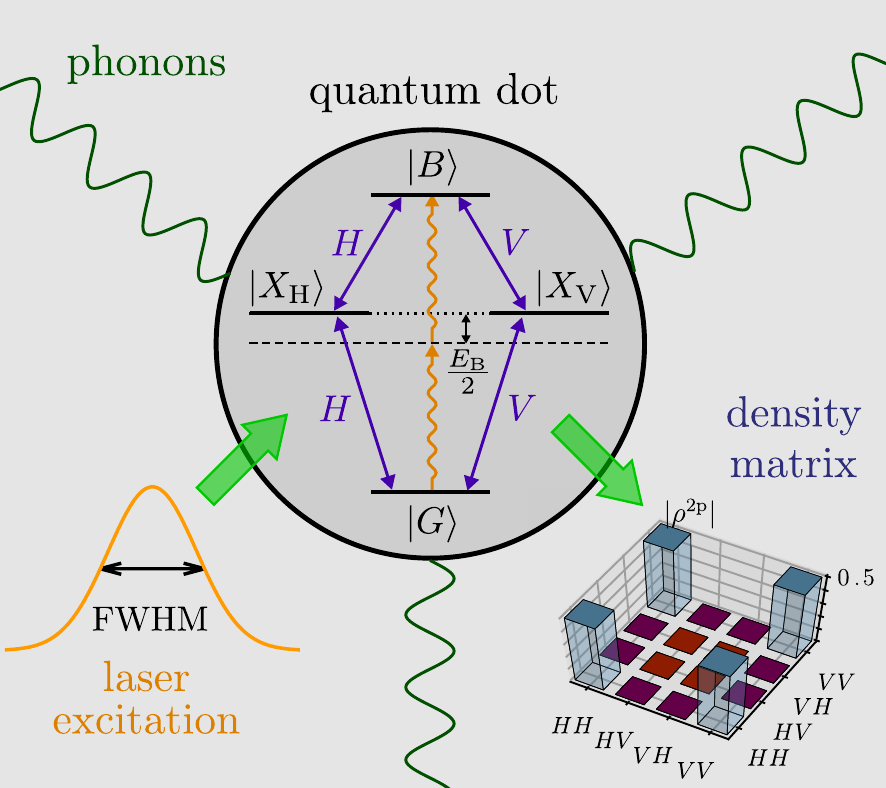}
\caption{
Sketch of the considered system. In the two-photon excitation scheme, a Gaussian laser pulse (left) with finite full-width-at-half-maximum (FWHM) is tuned to the two-photon resonance of a semiconductor quantum dot (center). Furthermore, due to its environment, the quantum dot is coupled to longitudinal acoustic phonons. The excited biexciton decays radiatively, following the biexciton-exciton cascade. In the ideal situation, either two horizontally ($H$) or two vertically ($V$) polarized photons are emitted, resulting in the creation of the maximally entangled state $\KET{\Phi_+}$, as described by the ideal two-photon density matrix $\rp$ (right).
}
\label{fig:system}
\end{figure}

We consider a strongly-confining InGaAs QD coupled to a continuum of LA phonons that is optically excited using the TPE scheme, as schematically illustrated in Figure~\ref{fig:system}. The biexciton is excited with a Gaussian laser pulse that is tuned to the two-photon resonance. Due to the strong confinement, higher excited states can be neglected and the QD is modeled as a four-level system. In the ideal situation, the decay of the biexciton yields the maximally entangled state $\KET{\Phi_+}$ described by the ideal two-photon density matrix shown in the lower right corner of Figure~\ref{fig:system}. 

\subsection{Optically excited strongly-confining quantum dots}
\label{subsec:sc_QDs}

The electronic structure of the QD is modeled as a four-level system, comprising the electronic ground state $\KET{G}$, two exciton states $\KET{X\sbt{H/V}}$ that couple to horizontally  and vertically polarized light, respectively, and the biexciton state $\KET{B}$. In the frame co-rotating with the external laser frequency $\omega\sbt{L}$, the corresponding Hamiltonian is given by
\begin{equation}
\label{eq:H_QD}
\begin{split}
\hat{H}\sbt{QD} =& \left(\Delta\sbt{XL}+\frac{\delta}{2}\right)\PRO{X\sbt{H}} + \left(\Delta\sbt{XL}-\frac{\delta}{2}\right)\PRO{X\sbt{V}}\\
&+ \left(2\Delta\sbt{XL}-E\sbt{B}\right)\PRO{B} ,
\end{split}
\end{equation}
where the energy of the ground state is used as the zero of the energy scale. Here $\Delta\sbt{XL}:=\hbar\left(\omega\sbt{X}-\omega\sbt{L}\right)$ is the detuning between the mean exciton energy $\hbar\omega\sbt{X}$ and the excitation laser, $\delta$ is the FSS, and $E\sbt{B}$ denotes the biexciton binding energy.

In the TPE scheme, the biexciton is directly excited with a short laser pulse tuned to the two-photon transition energy, i.e., $\Delta\sbt{XL} = E\sbt{B}/2$. We assume a Gaussian pulse shape with envelope function
\begin{equation}
\label{eq:Omega(t)}
\Omega(t) = \sqrt{\frac{4\ln(2)}{\pi}}\frac{\Theta}{\FWHM}\exp\left[ -4\ln(2)\left( \frac{t-t\sbt{L}}{\FWHM} \right)^2 \right] ,
\end{equation}
where the pulse duration is characterized by the full-width-at-half-maximum (FWHM) $\FWHM$. In typical experiments, the pulse duration is around 10~ps \cite{Huber_PRL_2018,Liu2019,Christian_Schimpf_Crypto_2021}. Note that the FWHM of the corresponding intensity profile $I(t)\propto\Omega^2(t)$ is $\FWHM/\sqrt{2}$. $t\sbt{L}$ denotes the time of the pulse maximum and  $\Theta$ is the (one-photon) pulse area. In the rotating wave approximation, the interaction between the QD and the laser pulse is described by the Hamiltonian
\begin{equation}
\label{eq:H_L}
\hat{H}\sbt{L} = \Omega(t)\left( \hat{\sigma}\sbt{L} + \hat{\sigma}\sbt{L}^\dagger \right) ,
\end{equation}
where the linear polarization of the laser is encoded in the coefficients $\alpha\sbt{H/V}\in\mathbb{R}$ in the operator
\begin{equation}
\hat{\sigma}\sbt{L} = \alpha\sbt{H} \sH + \alpha\sbt{V} \sV .
\end{equation}
Here the operator $\hat{\sigma}\sbt{H/V} := \OP{G}{X\sbt{H/V}} + \OP{X\sbt{H/V}}{B}$ describes the QD transitions coupled to horizontally/vertically polarized light. In this work, two different (linear) laser polarizations are considered: (i) horizontal laser polarization, i.e., $\alpha\sbt{H}=1$, $\alpha\sbt{V}=0$, and (ii) diagonal laser polarization, i.e., $\alpha\sbt{H}=\alpha\sbt{V}=1/\sqrt{2}$, with equal components in $H$ and $V$ direction.

The transitions between QD states that result in photon emission are modeled as radiative decay processes. They are described by Lindblad operators \cite{Lindblad:1976}
\begin{equation}
\label{eq:def_LB}
\LB{\hat{O}}{\Gamma}\hat{\rho} = \frac{\Gamma}{2}\left( 2\hat{O}\hat{\rho}\hat{O}^\dagger - \hat{O}^\dagger\hat{O}\hat{\rho} - \hat{\rho}\hat{O}^\dagger\hat{O} \right)
\end{equation}
with the corresponding transition operator $\hat{O}$ and rate $\Gamma$, that act on the statistical operator of the system $\hat{\rho}$. Due to the optical selection rules, the possible radiative decay processes are described by the four operators: $\LB{\OP{G}{X\sbt{H}}}{\gamma\sbt{X}}$, $\LB{\OP{G}{X\sbt{V}}}{\gamma\sbt{X}}$, $\LB{\OP{X\sbt{H}}{B}}{\frac{\gamma\sbt{B}}{2}}$, and $\LB{\OP{X\sbt{V}}{B}}{\frac{\gamma\sbt{B}}{2}}$. Here, $\gamma\sbt{B}$ ($\gamma\sbt{X}$) is the radiative decay rate of the biexciton (an exciton) state.

In strongly-confining InGaAs QDs, the most important electron-phonon interaction at low temperatures
is the deformation potential coupling to a continuum of
LA phonons \cite{Besombes:2001,Krummheuer:2002,Krummheuer2005,Uebersichtsartikel_2019}. This coupling is of the pure dephasing type and is described by the Hamiltonian
\begin{equation}
\label{eq:H_Ph}
\hat{H}\sbt{Ph} = \hbar\sum\limits_\bq \omega_\bq \hat{b}_\bq^\dagger\hat{b}_\bq + \hbar\sum\limits_{\chi,\bq} n_\chi \left( \gamma_\bq\spt{X}\hat{b}_\bq^\dagger + {\gamma_\bq\spt{X}}^\ast\hat{b}_\bq \right) \PRO{\chi} ,
\end{equation}
where the bosonic operator $\hat{b}_\bq^\dagger$ creates a phonon
in mode $\bq$ with energy $\hbar\omega_\bq$, $n_\chi = \left\lbrace
0,1,1,2\right\rbrace$ denotes the number of excitons present in the QD state
$\KET{\chi}=\left\lbrace
\KET{G},\KET{X\sbt{H}},\KET{X\sbt{V}},\KET{B}\right\rbrace$, and
$\gamma_\bq\spt{X}$ is the exciton-phonon coupling strength.
Here, we have classified the above exciton-phonon coupling as being of {\em pure dephasing type}
since the coupling is only to the occupations of the
states $|\chi\rangle$ and not to transitions between them.
It should be noted that combining such a pure dephasing type coupling with other interactions, the pure dephasing
coupling may also affect the dynamics of the occupations as is manifested, e.g., in the damping of
Rabi oscillations \cite{Knorr2003,pawel04,RabiRevival,McCutcheonPI}.

The dynamics of the statistical operator $\hat{\rho}$ of the complete system is governed by the Liouville-von Neumann equation
\begin{eqnarray}
\label{eq:LvN_Eq}
\ddt \hat{\rho} &=& - \frac{i}{\hbar} \left[ \hat{H},\hat{\rho} \right] + \sum\limits_{\ell=H,V} \left\lbrace \LB{\OP{G}{X_\ell}}{\gamma\sbt{X}} + \LB{\OP{X_\ell}{B}}{\frac{\gamma\sbt{B}}{2}} \right\rbrace \hat{\rho} \nonumber \\
 \\
\hat{H} &=& \hat{H}\sbt{QD} + \hat{H}\sbt{L} + \hat{H}\sbt{Ph} ,
\end{eqnarray}
where $[ \hat{A},\hat{B} ]$ is the commutator of two operators
$\hat{A}$ and $\hat{B}$. Employing a numerically complete real-time
path-integral method detailed in Refs.~\onlinecite{Makri_Theory,Makri_Numerics,PI_realtime2011,
PI_nonHamil2016,PI_cQED}, the dynamics of the reduced QD density matrix
$\hat{\bar{\rho}} = \Tr{Ph}{\hat{\rho}}$ is obtained. In the path-integral
algorithm, the coupling to LA phonons enters via the phonon spectral density
$J(\omega) = \sum\limits_\bq {\ABS{\gamma_\bq\spt{X}}}^2
\delta(\omega-\omega_\bq)$. An explicit expression for this quantity can be
found in Ref.~\onlinecite{PI_nonHamil2016} and the necessary GaAs material
parameters are taken from Ref.~\onlinecite{Krummheuer2005}.
In this work, we
consider a spherically symmetric InGaAs QD with a harmonic confinement and
electron (hole) confinement length $a_e = 3$~nm ($a_h = a_e/1.15$). Furthermore,
we use realistic parameters for the remaining QD properties, listed in
Table~\ref{tab:Fixed_Parameters}. For our numerical simulations, we assume that
the phonons are initially in a thermal equilibrium at temperature $T$, while the
QD is in its ground state $\KET{G}$. Note that for each simulation the optimal
pulse area $\Theta$ is determined numerically by optimizing for the maximum
biexciton occupation after the pulse.

The numerical completeness of our simulation means that, in order to evaluate the dynamical variables of interest,
no approximations to the model are made. We are neither using an expansion with respect to a presumably
small parameter such as the exciton-phonon or exciton-laser coupling nor do we neglect memory effects as done by
Markovian approaches. In particular, the memory functions representing the phonon-induced memory can be evaluated
beforehand \cite{PI_realtime2011,PI_cQED} such that the depth of the memory is known before starting the
path-integral simulation. Numerical errors arise in our simulation only due to discretization errors that are
easily checked for convergence.

\begin{table}
\centering
\caption{Realistic quantum dot parameters.}
\label{tab:Fixed_Parameters}
\begin{ruledtabular}
\begin{tabular}{l c c}
Parameter & & Value\\
\hline
Biexciton binding energy & $E\sbt{B}$ & 4~meV \\
Exciton-laser detuning & $\Delta\sbt{XL}$ & $E\sbt{B}/2=2$~meV \\
Radiative decay rate exciton & $\gamma\sbt{X}$ & 0.005~$\mathrm{ps^{-1}}$\\
Radiative decay rate biexciton & $\gamma\sbt{B}$ & $2\gamma\sbt{X}=0.010$~$\mathrm{ps^{-1}}$
\end{tabular}
\end{ruledtabular}
\end{table} 

\subsection{Reconstructed photonic density matrix and concurrence}
\label{subsec:2pdm&concurrence}

State-of-the-art experiments employ quantum state tomography \cite{QuantumStateTomography} to reconstruct the emitted two-photon state. This measurement scheme is based on polarization-resolved two-time coincidence measurements. The obtained signals are proportional to two-time correlation functions of photon operators. Because the sources of the emitted photons are electronic transitions between the QD states, i.e., the radiative decay, the two-time correlation function can be calculated using QD state transition operators as
\begin{equation}
\label{eq:def_G2}
\Gj(t,\tau) = \left\langle \hat{\sigma}_j^\dagger(t) \hat{\sigma}_k^\dagger(t+\tau) \hat{\sigma}_m(t+\tau) \hat{\sigma}_\ell(t) \right\rangle ,
\end{equation}
where $\left\lbrace j,k,\ell,m\right\rbrace\in\left\lbrace H,V \right\rbrace$. Here, $t$ denotes the time of the first photon detection event, while $\tau$ is the delay time until a subsequent second one occurs.

In experiments, one always measures quantities averaged over finite real time and delay time windows. Usually, the average interval for the real time $T\sbt{av}$ is chosen such that it encompasses the complete decay process. However, different subsets of emitted photons can be selected choosing different delay time windows $\tau\sbt{av}$ \cite{Different-Concurrences:18}. While a shorter delay time window is typically advantageous for the degree of entanglement \cite{StevensonPRL2008,Different-Concurrences:18,munoz15}, the corresponding photon yield is reduced.
Since, however, in applications often a high yield is mandatory,
all created photons should be utilized.
Thus, we aim for the maximum photon yield and consider all photon emission events.
Consequently, we calculate the two-photon density matrix $\rp$ as
\begin{eqnarray}
\label{eq:def_2pdm}
\rpj &=& \frac{\Gaj}{\Tr{}{\Ga}} , \\
\label{eq:def_G2av}
\Gaj &=& \lim_{T\sbt{av},\tau\sbt{av}\rightarrow \infty} \int\limits_0^{T\sbt{av}} \dt \int\limits_0^{\tau\sbt{av}} \mathrm{d}\tau \,\Gj(t,\tau) ,
\end{eqnarray}
where both time windows are taken in the limit of infinity.
Details on the numerical evaluation of two-time correlation functions within the path-integral formalism can be found in Ref.~\onlinecite{multi-time}.
It is important to note that our approach avoids using the quantum regression theorem (QRT).
The QRT is the standard tool for evaluating two-time correlation functions. However, its derivation relies on the assumption
that all environment influences are Markovian \cite{Carmichael:93,vma-238}, which is not true for phonons. Using the QRT naively in the bare state basis
yields emission spectra where the phonon sidebands appear energetically on the wrong side \cite{McCutcheon16}.
Working in a polaron transformed frame by using the polaron master equations (PMEs) the correct energetic position of the sidebands is recovered \cite{Roy-Choudhury2015,Iles-Smith2017,Reigue2017}.
However, even an advanced scheme like the PME,
which works very well for single-time density matrices
and captures many of the pertinent non-Markovian features of emission spectra,
turns out to yield noticeable errors in evaluating two-time correlation functions needed, e.g., for the indistinguishability \cite{vma-238-Erratum}.
Using a numerically complete scheme we are on the safe side to obtain correct values for the two-time correlation functions required for our present study.

The degree of entanglement associated with a reconstructed two-photon density
matrix is quantified using the concurrence, a well-established entanglement
measure that has a one-to-one correspondence to the entanglement of
formation \cite{Wootters1998,Wootters:2001}. In the basis
$\left\lbrace\KET{HH},\KET{HV},\KET{VH},\KET{VV}\right\rbrace$, the concurrence
$C$ is obtained directly from the two-photon density matrix
\begin{equation}
\label{eq:def_Con}
C = \max\left\lbrace0,\sqrt{\lambda_1}-\sqrt{\lambda_2}-\sqrt{\lambda_3}-\sqrt{\lambda_4}\right\rbrace ,
\end{equation}
where $\lambda_{j}\geq\lambda_{j+1}$ are the (real and positive) eigenvalues of the 4$\times$4-matrix
\begin{equation}
\label{eq:matrix_M}
M = \rp \, \Sigma \, (\rp)^\ast \, \Sigma \,.
\end{equation}
Here, $\Sigma$ is the anti-diagonal matrix with elements $\left\lbrace -1,1,1,-1\right\rbrace$ and $(\rp)^\ast$ denotes the complex conjugated two-photon density matrix.

\section{Phonon influence during two-photon excitation}
\label{sec:results}

The focus of this work is to investigate the combined impact of the TPE scheme and the pure dephasing-type coupling to LA phonons on the degree of entanglement of photon pairs emitted from strongly-confining QDs. First, the influence of phonons on the two-photon state is analyzed for a typical fixed pulse duration. Afterward, we study the combined impact for varying lengths of the excitation pulse.

\subsection{Temperature-dependent degradation of the entanglement}
\label{subsec:Con_vs_T_at_10ps}

\begin{figure*}[t]
\centering
\includegraphics[width=\textwidth]{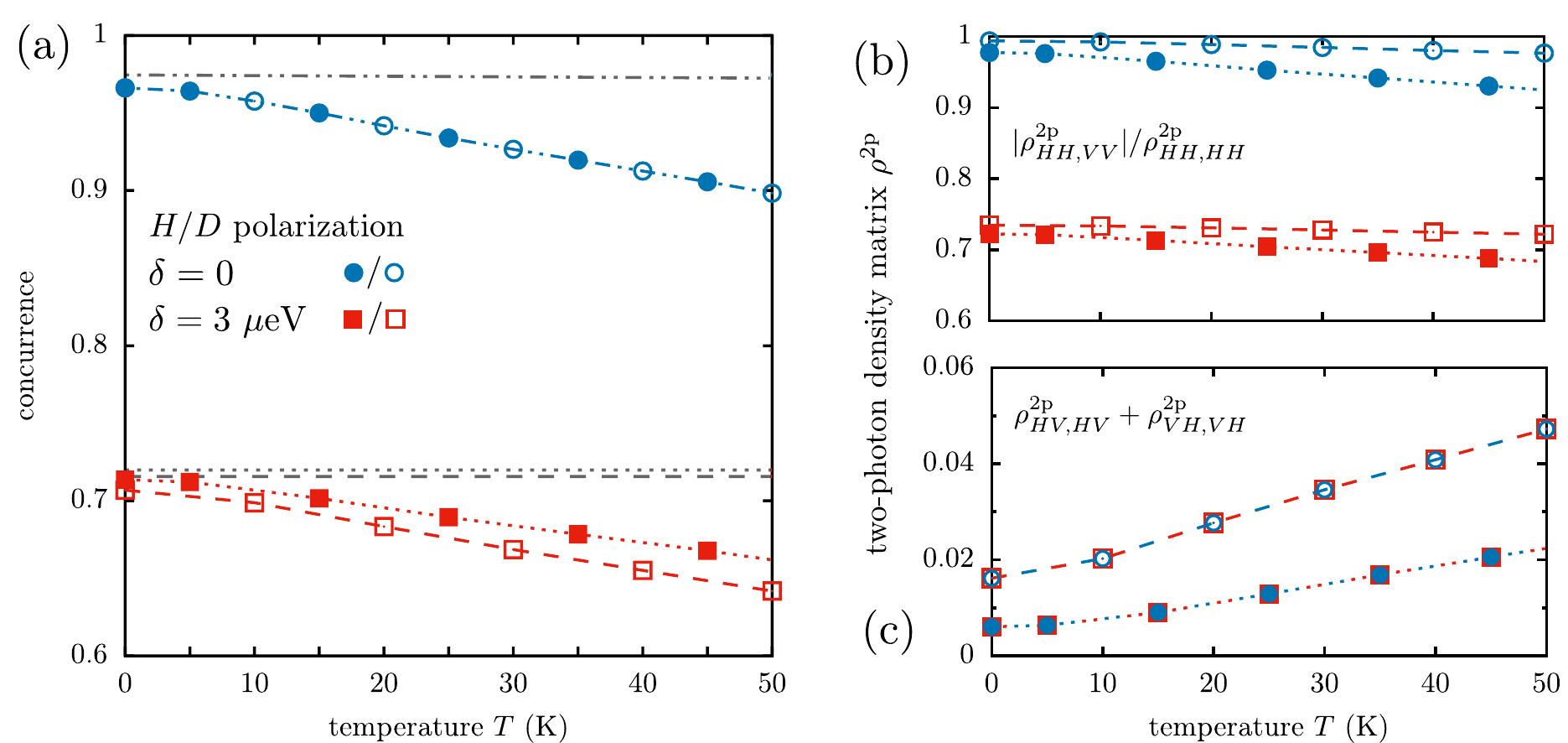}
\caption{ (a) Concurrence as a function of temperature for a typical pulse
duration $\FWHM=10$~ps. Two different laser polarizations [horizontal ($H$):
filled symbols and diagonal ($D$): open symbols] as well as two fine-structure
splittings $\delta=0$ (blue circles) and 3~$\mathrm{\mu}$eV (red squares) are
considered. The symbols represent numerical results and dotted (dashed) lines
indicate a guide to the eye for horizontal (diagonal) laser polarization. The
corresponding phonon-free results are indicated by gray horizontal lines with
the same dash type. For $\delta=0$, the results for $H$ and $D$ polarization are
exactly the same. (b) Ratio between the two-photon density matrix elements
$\ABS{\rpc}$ and $\rpH$, and (c) sum of the elements $\rpHV$ and $\rpVH$ for the
calculations presented in panel (a).  }
\label{fig:Con_vs_T}
\end{figure*}

We consider a typical pulse duration of $\FWHM = 10$~ps \cite{Huber_PRL_2018,Liu2019,Christian_Schimpf_Crypto_2021} and investigate the temperature dependence of the obtained concurrence. Figure~\ref{fig:Con_vs_T}(a) displays the concurrence as a function of temperature $T$ for horizontal ($H$, filled symbols) and diagonal ($D$, open symbols) laser polarization. A vanishing FSS (blue) as well as a finite FSS of $\delta=3$~$\mathrm{\mu}$eV (red) is considered. In all cases, the degree of entanglement decreases monotonically with rising temperature, revealing a phonon-induced degradation of the entanglement. While the obtained concurrence is independent of the laser polarization for a vanishing FSS, the difference $\Delta C$ between the concurrence obtained for horizontal and diagonal laser polarization increases with temperature when the FSS is finite. Although the concurrence comes close to the corresponding phonon-free result (gray horizontal lines) in the limit $T\rightarrow 0$, the latter is not recovered, even for $T=0$. 

In order to understand the phonon influence, we briefly recapitulate the results
for an ideal four-level quantum emitter, i.e., the phonon-free case, discussed
in Ref.~\onlinecite{Seidelmann_PRL_2022}. It was found that the degree of
entanglement is limited due to a laser-induced which-path information introduced
during the excitation pulse; cf. gray horizontal lines in
Figure~\ref{fig:Con_vs_T}(a). Because there is a finite probability that the
biexciton decays into an exciton state already during the pulse, this which-path
information impacts the emitted two-photon state and reduces the
concurrence. Note that reexcitation is hardly a factor in the phonon-free case
since the one-photon process bringing the QD from the exciton state back into
the biexciton is strongly detuned from the laser energy. In the polarization
basis $\left\lbrace H,V\right\rbrace$, the laser-induced which-path information
is interpreted differently, depending on the chosen laser polarization. In the
case of the horizontal laser polarization, an energy splitting on the order of
$E\sbt{S}\sim\hbar\pi/\FWHM$ is introduced between the two exciton states
$\KET{X\sbt{H}}$ and $\KET{X\sbt{V}}$ due to the AC Stark effect. Similar to the
FSS, the splitting $E\sbt{S}$ results in dephasing and a reduced coherence
$\rpc$ in the two-photon density matrix. On the other hand, a diagonal laser
polarization introduces an effective coupling between these exciton states,
creating undesired photon states $\KET{HV}$ and $\KET{VH}$. Because all linear
laser polarizations are equivalent, in the case of a vanishing FSS, the
resulting concurrence is exactly the same for both laser polarizations. However,
if the FSS is finite, using a horizontal laser polarization gives a slightly
higher concurrence.

Similar to the phonon-free case, the phonon effects that reduce the degree of
entanglement depend on the laser polarization. We start the analysis with the
case of the horizontal laser polarization. In this situation, the loss of
entanglement can be traced back to two phonon-related effects. The laser couples
only to QD transitions involving $\KET{X\sbt{H}}$ and, thus, introduces an
effective splitting $E\sbt{S}$ between the exciton states. Because the exciton
states are energetically split during the pulse, phonons can reduce the
concurrence by increasing dephasing \cite{Seidelmann2019}. Due to the effective
splitting the two decay paths are associated with different transition
energies. Thus, phonons enlarge the which-path information as energies of
involved phonons depend on the decay path. This enlarged which-path information
results in phonon-enhanced dephasing and a reduced coherence between the
two-photon states $\KET{HH}$ and $\KET{VV}$; cf. filled symbols in
Figure~\ref{fig:Con_vs_T}(b). When the temperature increases, more phonons
become thermally excited. Thus, the phonon-induced dephasing is the stronger the
higher the temperature, and the corresponding coherence $\rpc$ and concurrence
decrease with rising $T$.

In addition, the interaction with LA phonons also increases the reexcitation
probability. Phonon absorption and emission processes can assist the
off-resonant single-photon processes by compensating the energy mismatch between
the QD transition energies and the energy of the
laser \cite{Seidelmann2019}. Thus, when the biexciton decays into the exciton
state $\KET{X\sbt{H}}$, under emission of a $H$ polarized photon, already during
the pulse, a reexcitation of the biexciton state induced by the same laser
pulse becomes more likely since a phonon with the energy covering the detuning
can be emitted. Because the biexciton can now again decay into either exciton
state, the undesired state $\KET{HV}$ can be created, which also reduces the
degree of entanglement. Also the probabilities of phonon absorption and emission
processes increase with rising temperature. Consequently, phonon-assisted
reexcitation processes are more likely to occur at higher temperatures. This
effect can be seen in Figure~\ref{fig:Con_vs_T}(c), which shows an increasing
probability to find a detrimental two-photon state $\rpHV+\rpVH$ with rising
temperature; cf. filled symbols. Note that for a horizontal laser polarization,
$\KET{X\sbt{V}}$ is decoupled from the laser excitation. Thus, two-photon states
$\KET{VH}$ are insignificant, i.e.,  $\rpHV\gg\rpVH$ (not shown). Because both
phonon-related effects that degrade the entanglement -- the phonon-induced
dephasing and the phonon-assisted reexcitation -- increase with temperature, the
resulting concurrence decreases monotonically as a function of $T$. Note that
the phonon-free result (gray horizontal line) is not recovered for $T=0$ because
phonon emission processes are still possible. Furthermore, because typical FSSs
are much smaller than the energy mismatch between the laser and QD transition
energies, i.e., $\delta\ll E\sbt{B}/2$, the energy of phonons that needs to be
absorbed or emitted hardly changes. Thus, the associated reexcitation
probability, and hence $\rpHV+\rpVH$, remain the same when considering a finite
FSS $\delta=3$~$\mathrm{\mu}$eV; cf. red and blue symbols in
Figure~\ref{fig:Con_vs_T}(c).

Next, we turn to a diagonal laser polarization. In this situation, the phonon-related mechanism that causes the degradation of entanglement with rising $T$ is slightly different. Essentially, only one of the two effects discussed above, i.e., phonon-assisted reexcitation, is relevant here. For a diagonal laser polarization, the TPE scheme introduces an effective coupling between the exciton states $\KET{X\sbt{H}}$ and $\KET{X\sbt{V}}$, rather than an energy splitting. Therefore, the phonon-induced dephasing is marginal and the relative coherence $\rpc$ is hardly reduced; cf. open symbols in Figure~\ref{fig:Con_vs_T}(b). On the other hand, both exciton states are coupled to the laser excitation. Consequently, phonon-assisted reexcitation leads to a much higher probability to obtain detrimental two-photon states; cf. open symbols in Figure~\ref{fig:Con_vs_T}(c). Note that for a diagonal laser polarization the system is symmetric with respect to $H$ and $V$ polarization. Thus, the probability to find mixed two-photon states $\KET{HV}$ and $\KET{VH}$ is equal, i.e., $\rpHV=\rpVH$ (not shown).

For a vanishing FSS, no linear polarization basis is distinguished, even when the coupling to phonons is present. Thus, from a physical point of view, the concurrence has to be independent of the chosen (linear) laser polarization. Indeed, the concurrence for horizontal and diagonal laser polarization is exactly the same; cf. blue symbols in Figure~\ref{fig:Con_vs_T}(a). But, for a fixed polarization basis, the weights of the two phonon effects depend on the chosen laser polarization. Thus, the phonon impact on elements of the two-photon density matrix differs for horizontal and diagonal laser polarization; cf. panels (b) and (c). However, for vanishing FSS, both cases describe the same physical situation, just from a different perspective. Consequently, the two different two-photon density matrices obtained for horizontal and diagonal laser polarization are linked by a simple basis transformation. This means, when one transforms the polarization basis from $H$ and $V$ polarization to the diagonal $D=(H+V)/\sqrt{2}$ and anti-diagonal $A=(H-V)/\sqrt{2}$ basis, the two different density matrices are precisely converted into each other. 

However, when the FSS is finite the basis $\lbrace H,V\rbrace$ is distinguished and the concurrence depends on the chosen laser polarization. As in the phonon-free case, the horizontal laser polarization gives a higher concurrence compared to a diagonal one. Furthermore, the difference in concurrence $\Delta C$ is enlarged by the interaction with LA phonons and increases with temperature.

\subsection{Increasing phonon impact for longer pulses}
\label{subsec:Con_vs_FWHM}

After the origins for the observed temperature-dependent degradation of entanglement have been identified to be phonon-induced dephasing and phonon-assisted reexcitation, we now investigate the concurrence as a function of the pulse duration. Although the laser-induced splitting or coupling between the exciton states decreases with rising pulse duration (because the pulse peak intensity is decreased when the pulse area is kept fixed), more emission events are impacted by the laser-induced effects when the pulse becomes longer. Consequently, already in the phonon-free case, the achievable degree of entanglement drops with rising FWHM \cite{Seidelmann_PRL_2022}. In addition, also the detrimental phonon-related effects have more time to impact the system when the pulse duration is increased. Consequently, at a given temperature, the concurrence should decrease for longer pulses. Furthermore, the drop in concurrence with increasing temperature should be the stronger the longer the pulse duration is.

\begin{figure}[t]
\centering
\includegraphics[width=\columnwidth]{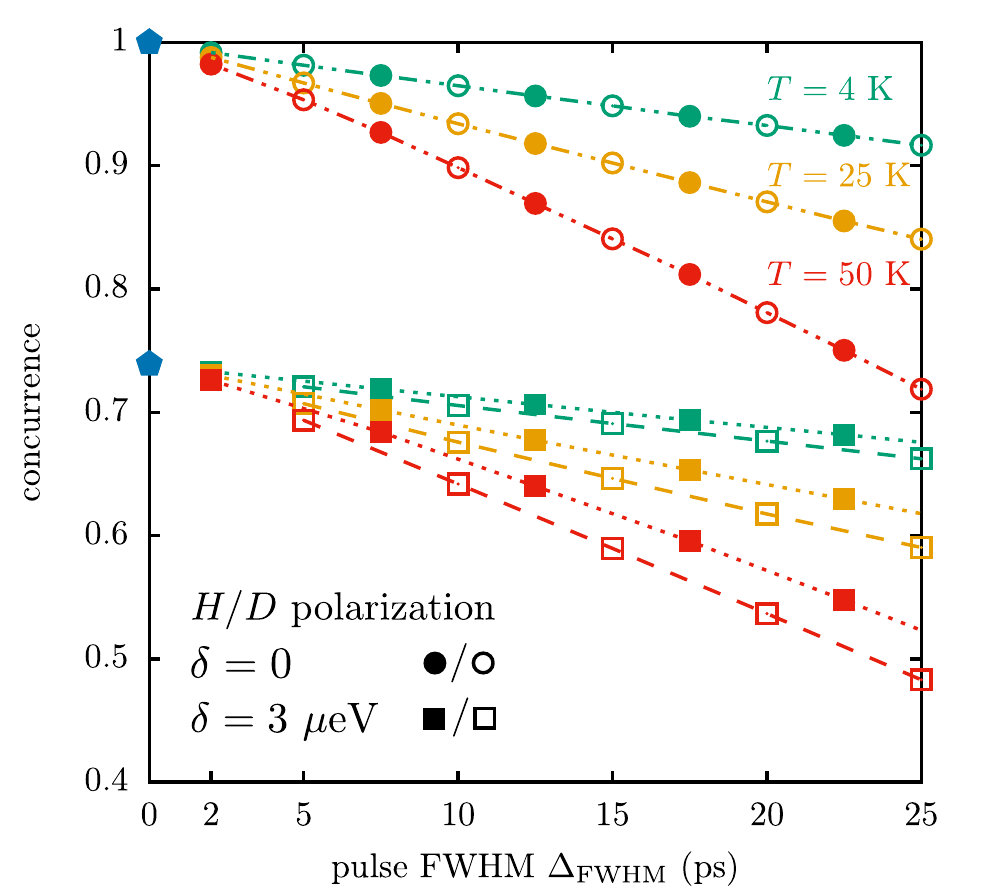}
\caption{
Concurrence in dependence of the pulse duration, characterized by its FWHM, for two laser polarizations [horizontal ($H$): filled symbols and diagonal ($D$): open symbols] and two fine-structure splittings: $\delta=0$ (circles) and 3 $\mathrm{\mu}$eV (squares). Three different temperatures are considered: $T=4$~K (green), 25~K (orange), and 50~K (red). The symbols represent numerical results and dotted (dashed) lines indicate a guide to the eye for horizontal (diagonal) laser polarization. For $\delta=0$, the results for $H$ and $D$ polarization are exactly the same. Data points at $\FWHM=0$ (blue pentagons) represent phonon-free calculations for an initially prepared biexciton without optical excitation.
}
\label{fig:Con_vs_FWHM}
\end{figure}

The numerical results presented in Figure~\ref{fig:Con_vs_FWHM} confirm these expectations. For a given temperature, the concurrence decreases monotonically when the pulse duration is increased. Furthermore, the slope of this decrease steepens with rising FWHM. Thus, both phonon-related mechanisms, indeed, become more relevant for longer pulses. For a vanishing FSS, all linear laser polarizations are equivalent, resulting again in exactly the same concurrence for horizontal and diagonal polarization. Because the phonon-related effects are more important at higher temperatures and/or longer pulse durations, the difference in concurrence between the two laser polarizations found for a finite FSS $\delta=3$~$\mathrm{\mu}$eV increases as well.

In the limit of $\FWHM\rightarrow 0$ the combined, detrimental effect of TPE and LA phonons fades away, and the corresponding concurrence approaches the phonon-free result for an initially prepared biexciton, indicated by blue pentagons in Figure~\ref{fig:Con_vs_FWHM}. In this sense it is indeed the finite pulse duration that unlocks the phonon impact on the concurrence. However,  one has to keep in mind that the TPE scheme breaks down in this limit. For a typical GaAs-based QD, a pulse duration below $\lesssim 3$~ps results in a strongly reduced preparation probability of the biexciton, because the energy spectrum of the laser pulse begins to overlap with the exciton transition energies \cite{Stefanatos2020_Quantum_speed_limit}. Thus, below these pulse durations, the concurrence can only be enhanced if one is willing to accept a severe decline in the photon yield. Note that the reduced probability to excite the biexciton does not reduce the concurrence, because the coincidence measurements only consider cases where two photons are emitted.

\section{Conclusion}
\label{sec:conclusion}

We have investigated the combined impact of TPE and the pure dephasing-type
coupling to LA phonons on the achievable entanglement of photon pairs emitted
from strongly-confining QDs. It is found that the interaction with phonons
reduces the degree of entanglement, as measured by the concurrence, further
below the recently established limitation due to the TPE scheme itself. At a
fixed pulse duration, the combined impact results in a monotonic decrease of the
concurrence when the temperature increases. This phonon-induced degradation
of the entanglement does not vanish in the limit of vanishing temperature.
Depending on the chosen laser
polarization, the origin of this temperature-dependent degradation is found to
be a combination of phonon-induced dephasing and phonon-assisted one-photon
processes that increase the probability to find the mixed states $\KET{HV}$ and
$\KET{VH}$ after reexcitation of the biexciton.

Because phonons have more time to influence the system for longer pulses the degree of entanglement decreases with rising temperature and/or pulse duration, even in the absence of an excitonic fine-structure splitting. Furthermore, when the fine-structure splitting is finite, the coupling to phonons enlarges the difference in concurrence obtained for different laser polarizations that is already present in the phonon-free case. Although the combined, detrimental impact of TPE and LA phonons disappears in the limit of zero pulse duration and vanishing FSS, the TPE scheme in QDs typically breaks down below a FWHM of $\FWHM\lesssim 3$~ps. Hence, in realistic situations, one will always observe a temperature-dependent degradation of entanglement for photon pairs emitted by a strongly-confining QD.

Finally, we would like to mention that a temperature-dependent drop of the concurrence has recently also been observed in the case of vanishing FSS for QDs that are not in the strong-confinement limit  \cite{Lehner_ExcitedStates_2022}.
For these larger quantum dots, the pure dephasing-like coupling to phonons becomes negligible because the effective coupling strength decreases with rising QD size \cite{Glaessl_QDsize_2011,Lueker_QDgeom_2017}.
However, the phonon-mediated coupling to higher energetic states has been identified in that study as the origin of the reduction in concurrence. For strongly-confining QDs, we are dealing here with the reverse situation where couplings to higher energetic states are negligible but pure dephasing becomes important. Indeed, in our present study, we demonstrate a noticeable degradation of the concurrence even when higher states are completely absent. Interestingly, the temperature dependence observed in Ref.~\onlinecite{Lehner_ExcitedStates_2022} differs from our present result: there the concurrence stays on a plateau for lower temperatures before it drops much more steeply than found in the present paper. Indeed, while we find in Fig.~\ref{fig:Con_vs_T}(a) for $\delta=0$ and $T=50$~K a concurrence that is still $\sim$0.9, for the larger QDs in Ref.~\onlinecite{Lehner_ExcitedStates_2022} the corresponding value was almost reduced to 0.4. Thus, we see a clear advantage of strongly-confining QDs at elevated temperatures.  On the other hand, strongly-confining QDs are usually characterized by smaller optical transition rates, which make these systems more vulnerable to residual FSS and other noise sources. A holistic treatment of QD excitonic structure and excitation scheme, and a detailed knowledge of dephasing and reexcitation sources will thus be needed to create QDs capable of generating maximally entangled photons at elevated temperatures.

\section*{Acknowledgments}
T. K. B. and D. E. R. acknowledge support by the Deutsche Forschungsgemeinschaft (DFG) via the project No.~428026575.
A. R. acknowledges financial support by the Austrian Science Fund (FWF) via the Research Group FG5, P~29603, P~30459, I~4320, I~4380, I~3762, the European Union's Horizon 2020 research and innovation program under Grant Agreements No.~899814 (Qurope) and No.~871130 (Ascent+), the QuantERA II Programme that has received funding from the European Union's Horizon 2020 research and innovation program under Grant Agreement No.~101017733 via the project QD-E-QKD and the FFG Grant No.~891366, the Linz Institute of Technology (LIT), and the LIT Secure and Correct Systems Lab, supported by the State of Upper Austria.
We are further grateful for support by the Deutsche Forschungsgemeinschaft (DFG, German Research Foundation) via the project No.~419036043.

\bibliography{PIbib}

\end{document}